# Diversity Limits of Compact Broadband Multi-Antenna Systems

Pawandeep S. Taluja, *Member, IEEE,* and Brian L. Hughes, *Member, IEEE*

***Abstract*—In order to support multiple antennas on compact wireless devices, transceivers are often designed with matching networks that compensate for mutual coupling. Some works have suggested that when optimal matching is applied to such a system, performance at the center frequency can be improved at the expense of an apparent reduction in the system bandwidth. This paper addresses the question of how coupling impacts bandwidth in the context of circular arrays. It will be shown that mutual coupling creates eigen-modes (virtual antennas) with diverse frequency responses, using the standard matching techniques. We shall also demonstrate how common communications techniques such as Diversity-OFDM would need to be optimized in order to compensate for these effects.**

***Index Terms*—multiple antennas, mutual coupling, broadband matching, MIMO, OFDM.**

## I. Introduction

MULTIPLE-antenna systems have been shown to alleviate the problem of signal fading, and promise high spectral efficiencies in wireless propagation environments rich in multipath [1], [2]. However, most of these advantages from multiple-input multiple-output (MIMO) systems can only be realized for large antenna spacings. As the form-factor of current handheld and portable devices continues to shrink, and the demand for high data rate systems continues to grow, it is becoming increasingly necessary to deploy multiple antennas in a small space. Several wireless standards, including 4G LTE, IEEE 802.11n, provide support for multi-antenna devices such as mobile phones, laptops, tablets or access points – some of which offer a great fit for two dimensional arrays, popularly circular.

With closely spaced antennas, impairments such as fading correlation and mutual coupling become increasingly dominant. Transceivers designed for compact antenna arrays often employ special radio-frequency (RF) networks called matching networks that are embedded between the antenna array and the rest of the RF chain designed for optimized performance. Several studies have proposed optimal transceiver design for MIMO systems in the presence of mutual coupling by the use of multiport matching networks [3]–[8]. However, the focus of these studies has largely been narrowband systems; the effects of mutual coupling on the bandwidth of MIMO systems has received little attention. These results suggest that when optimal matching is applied to a system with strong mutual coupling, performance at the center frequency can be improved at the expense of an apparent reduction in the system bandwidth [9], [10]. This raises fundamental questions about the physical realizability of these networks and the bandwidth assumptions.

The study of matching networks for wideband systems falls under the purview of broadband matching theory – a field rich in multiport broadband matching network design, albeit, limited to systems not involving coupled sources or loads. However, there exist techniques in antenna/microwave circuit design that deal with coupled systems by use of orthogonal beam-forming networks [11], and apply matching to the decoupled ports of the antenna array.

In this paper, we investigate optimal transceiver design for compact arrays from a communication theory perspective. Specifically, we focus on the impact coupling has on the bandwidth of coupled circular arrays and derive optimal broadband matching networks. It will be shown that for broadband systems with uniform circular arrays, mutual coupling decomposes the coupled array with spectrally-identical spatial modes into spectrally non-identical eigen-modes – with differing bandwidths and resonant-frequencies. Similar results have appeared for specialized cases recently [12], [13]. These studies treat the problem in detail from a microwave theory standpoint, including many of the implementation aspects. Our findings and analysis help generalize how coupling impacts the RF bandwidth in the context of circular arrays. By combining concepts from antenna/microwave theory, Fano's broadband matching and Shannon's information theory, we present a unified communication-theoretic framework and evaluate the diversity limits of coupled broadband systems – utilizing orthogonal frequency division multiplexing (OFDM) – with varying antenna spacing.

The organization of the paper is as follows. Sec. II presents an overview of the basic microwave theory tools necessary to analyze the problem of coupled MIMO systems. Sec. III illustrates how coupling impacts the RF bandwidth of a compact circular array and introduces the concept of virtual antennas. It also discusses the applicability of Fano's broadband matching theory to characterize the optimal broadband matching network. Sec. IV supports these findings with numerical data. In Sec. V, we present a transceiver model for a broadband coupled array, and employ it in Sec. VI to develop a system model for

P. S. Taluja is with the Communications Systems Group at MaxLinear Inc., Carlsbad, CA, USA (email: pstaluja@ncsu.edu).

B. L. Hughes is with the Department of Electrical and Computer Engineering, North Carolina State University, Raleigh, NC, USA (email: blhughes@ncsu.edu).

This material is based upon work supported by the National Science Foundation under grant CCF-1018382. Portions of this paper were presented at the 2011 IEEE Global Communications Conference.

Diversity-OFDM in the presence of mutual coupling. Sec. VII presents results for the outage capacity as a function of antenna separation. We conclude by summarizing the main findings of the paper in Sec. VIII.

## II. Coupled Transceivers

As the antennas in an array are brought closer, currents flowing in one element induce voltage across the other. This is commonly referred to as mutual coupling. This phenomenon is usually modeled using the circuit (or network) representation of the antenna array. For an $N$ element antenna array, the currents $\mathbf{i}$ flowing through the antenna ports, and voltages $\mathbf{v}$ induced across them can be modeled by $N \times 1$ vectors and $N \times N$ matrices as:

$$\mathbf{v} = \mathbf{Z}_A \mathbf{i} + \mathbf{v}_o \ .$$

Here, $\mathbf{Z}_A$ denotes the $N \times N$ antenna array impedance matrix and $\mathbf{v}_o$ the $N \times 1$ open-circuit voltage induced by the incident electro-magnetic field (EM). The diagonal entries of $\mathbf{Z}_A$ denote self-impedance and the non-diagonal ones mutual-impedance, such that in the absence of coupling, the antenna impedance matrix is diagonal. It has been shown that optimal performance can be achieved by transforming the coupled antenna array to an uncoupled one by use of a $2N$-port impedance-transforming network called matching network $\mathbf{Z}_M$, inserted between the array and the rest of the transceiver [4], [7]. The choice of the matching network depends on the antenna separation (essentially the extent of coupling) and is usually chosen as a fixed passive network. The optimal networks are generally non-diagonal in nature, while the more practical, but sub-optimal, are diagonal. However, these studies have essentially focused on narrowband models and in order to incorporate bandwidth considerations, the analysis must be extended to using matching networks that perform well over a given bandwidth. As will be shown later, the system analysis of broadband systems with mutual coupling is eased by the use of scattering-parameter or S-matrix representation, instead of the impedance matrix.

To begin, we introduce the basic elements of broadband matching theory for a single antenna (or 2-port network) system. Consider a broadband 2-port network (shown in Fig. 1), with impedance matrix $\mathbf{Z}(s)$ terminated into reference impedances $z_1(s)$ and $z_2(s)$ on either side. Here, $s = \sigma + j\omega$ is the Laplace variable and $\omega = 2\pi f$ denotes frequency in radians/sec. For this simple (assuming reciprocal) 2-port network, the elements of S-matrix represent the reflection and transmission coefficients $\Gamma_1$, $\Gamma_2$ and $T$, respectively, as shown in Fig. 1:

$$\mathbf{S}(s) = \begin{bmatrix} \Gamma_1(s) & T(s) \\ T(s) & \Gamma_2(s) \end{bmatrix} \ .$$

It relates the voltages across and currents through the ports to the incident and reflected normalized *wave vectors* $\mathbf{a}$ and $\mathbf{b}$, respectively, via [14]

$$\mathbf{b}(s) = \mathbf{S}(s)\mathbf{a}(s) \ , \tag{1}$$

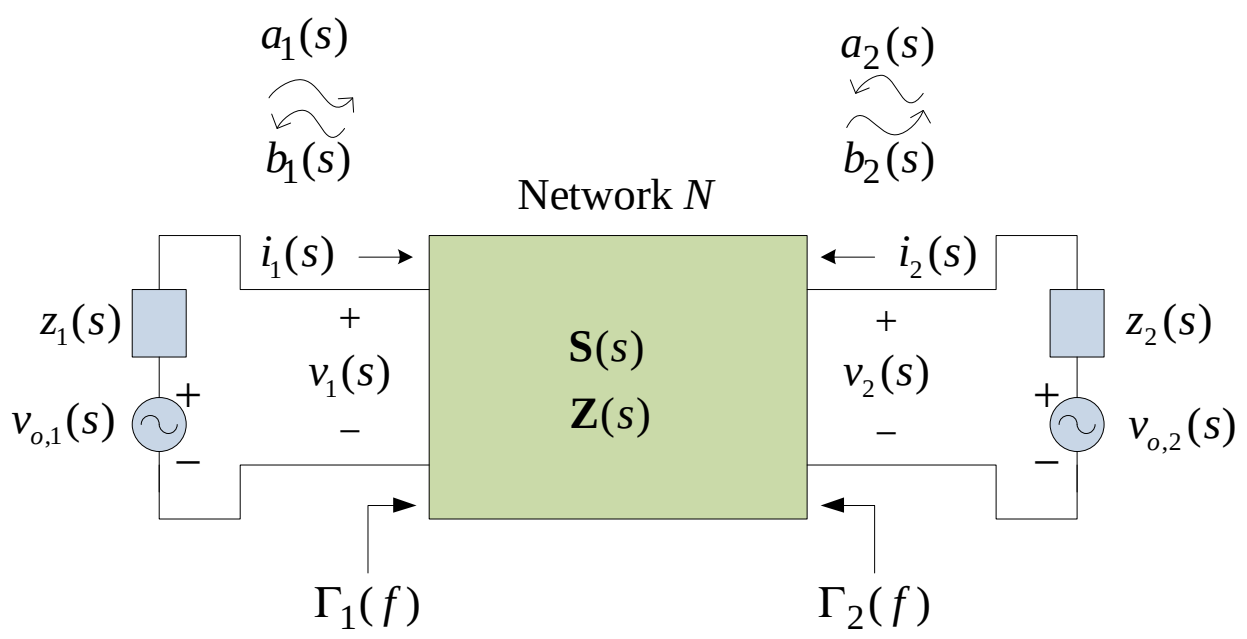

Fig. 1. Input and output wave vectors for a two-port network

where,

$$\mathbf{a}(s) = \begin{bmatrix} a_1(s) \\ a_2(s) \end{bmatrix} \ , \ \mathbf{b}(s) = \begin{bmatrix} b_1(s) \\ b_2(s) \end{bmatrix} \ .$$

For impedances $z_1(s) = z_2(s) = 1 \ \Omega$, the S-matrix is readily computed using the impedance matrix

$$\mathbf{S}(s) = (\mathbf{Z}(s) + \mathbf{I})^{-1}(\mathbf{Z}(s) - \mathbf{I}) \ , \tag{2}$$

where $\mathbf{I}$ denotes the identity matrix. For a lossless and reciprocal network, $\mathbf{S}$ satisfies

$$\mathbf{S}(j\omega)\mathbf{S}^H(j\omega) = \mathbf{I} \ , \quad \mathbf{S}(j\omega) = \mathbf{S}^T(j\omega) \ .$$

In the next section, we discuss a very special but practical class of planar antenna arrays – uniform circular arrays (UCA). We shall use the circulant nature of a UCA to ease the analysis and gain rich insights into the impact of coupling on the bandwidth of compact arrays.

## III. Virtual Antennas

We begin with characterization of the broadband antenna array S-matrix $\mathbf{S}_A$. In order to establish $\mathbf{S}_A$, it is convenient to look at the impedance matrix $\mathbf{Z}_A$. For example, an antenna array with $N = 2$ identical elements placed a distance $d$ apart,[1] has a symmetric impedance matrix of the form (cf. Fig. 2)

$$\mathbf{Z}_A(j\omega) = \begin{bmatrix} z_{11}(j\omega) & z_{12}(j\omega) \\ z_{12}(j\omega) & z_{11}(j\omega) \end{bmatrix} \ .$$

The symmetric nature of $\mathbf{Z}_A$ above enables us to express it in terms of its eigen-value decomposition (EVD)

$$\mathbf{Z}_A(j\omega) = \mathbf{Q}\mathbf{\Lambda}_A(j\omega)\mathbf{Q}^H \ , \tag{3}$$

where the set of unitary eigen-vectors is given by

$$\mathbf{Q} \ = \ \frac{1}{\sqrt{2}} \begin{bmatrix} 1 & 1 \\ 1 & -1 \end{bmatrix} \ ,$$

and the eigen-values by

$$\lambda(j\omega) = \{z_{11}(j\omega) + z_{12}(j\omega), \ z_{11}(j\omega) - z_{12}(j\omega)\} \ , \tag{4}$$

henceforth referred to as *eigen-impedances*. It is important to point out that the unitary transformation $\mathbf{Q}$ above is frequency-independent. This makes the analysis of the broadband matching problem applicable to coupled arrays, and the implementation of matching network considerably easier. As will be discussed later, the matching network can be realized as a cascade of a frequency-independent decoupling network followed by a diagonal broadband matching network.

[1] The antenna separation $d$ is specified in terms of $\lambda_c$ – wavelength corresponding to the center frequency $f_c$.

### A. Uniform Circular Arrays

A careful observation of the structure of $\mathbf{Z}_A$ and the unitary transformation $\mathbf{Q}$ reveals that it is straightforward to extend the entire theory to uniform circular arrays. A uniform circular array has a circulant impedance matrix. For a circulant $\mathbf{Z}_A$ with $N$ elements, the eigen-vectors for $\mathbf{Z}_A(j\omega) = \mathbf{Q}\mathbf{\Lambda}_A(j\omega)\mathbf{Q}^H$, are given by the columns of the unitary matrix

$$\mathbf{Q} = \frac{1}{\sqrt{N}} \begin{bmatrix} 1 & 1 & \dots & 1 \\ 1 & \alpha & \dots & \alpha^{N-1} \\ 1 & \alpha^2 & \dots & \alpha^{2(N-1)} \\ \vdots & \vdots & & \vdots \\ 1 & \alpha^{N-1} & \dots & \alpha^{(N-1)(N-1)} \end{bmatrix}$$

where $\alpha = e^{-2\pi j/N}$. The eigen-values ($\mathbf{\Lambda}_A$) are given by the discrete Fourier transform (DFT) of the first row of $\mathbf{Z}_A$.

For example, the spatial unitary transformation that decouples an $N = 3$ UCA, is given by

$$\mathbf{Q} = \frac{1}{\sqrt{3}} \begin{bmatrix} 1 & 1 & 1 \\ 1 & -\frac{1}{2} - j\frac{\sqrt{3}}{2} & -\frac{1}{2} + j\frac{\sqrt{3}}{2} \\ 1 & -\frac{1}{2} + j\frac{\sqrt{3}}{2} & -\frac{1}{2} - j\frac{\sqrt{3}}{2} \end{bmatrix}$$

and the eigen-impedances by

$$\lambda_1(j\omega) = z_{11}(j\omega) + 2z_{12}(j\omega) \ , \tag{5a}$$
$$\lambda_2(j\omega) = z_{11}(j\omega) - z_{12}(j\omega) \ , \tag{5b}$$
$$\lambda_3(j\omega) = \lambda_2(j\omega) \ . \tag{5c}$$

Similarly for $N = 4$ antennas, we have

$$\mathbf{Q} = \frac{1}{2} \begin{bmatrix} 1 & 1 & 1 & 1 \\ 1 & -j & -1 & j \\ 1 & -1 & 1 & -1 \\ 1 & j & -1 & -j \end{bmatrix}$$

and eigen-impedances as

$$\lambda_1(j\omega) = z_{11}(j\omega) + 2z_{12}(j\omega) + z_{13}(j\omega) \ , \tag{6a}$$
$$\lambda_2(j\omega) = z_{11}(j\omega) - z_{13}(j\omega) \ , \tag{6b}$$
$$\lambda_3(j\omega) = z_{11}(j\omega) - 2z_{12}(j\omega) + z_{13}(j\omega) \ , \tag{6c}$$
$$\lambda_4(j\omega) = \lambda_2(j\omega) \ . \tag{6d}$$

The unitary transformation $\mathbf{Q}$, is essentially an orthogonal beam-forming matrix implemented using RF networks called beam-formers [11]. These beam-formers are capable of producing $N$ spatially orthogonal beams, hence the operation represents a spatial DFT.[2] In terms of the antenna radiation patterns, this operation can be thought of as decomposing the composite array pattern into overlapping, but mutually orthogonal patterns at the output of the beam-former. For an illustration of such pattern decomposition, see [12], [13]. Similar analysis, however, can not be extended to uniform linear arrays for which the array impedance matrix is complex-symmetric.[3]

[2] One of the well known and widely used matrix for arrays with $2^n$ number of antennas is the Butler matrix [15], representing spatial FFT operations. However, our case necessitates a more general decoupling matrix.

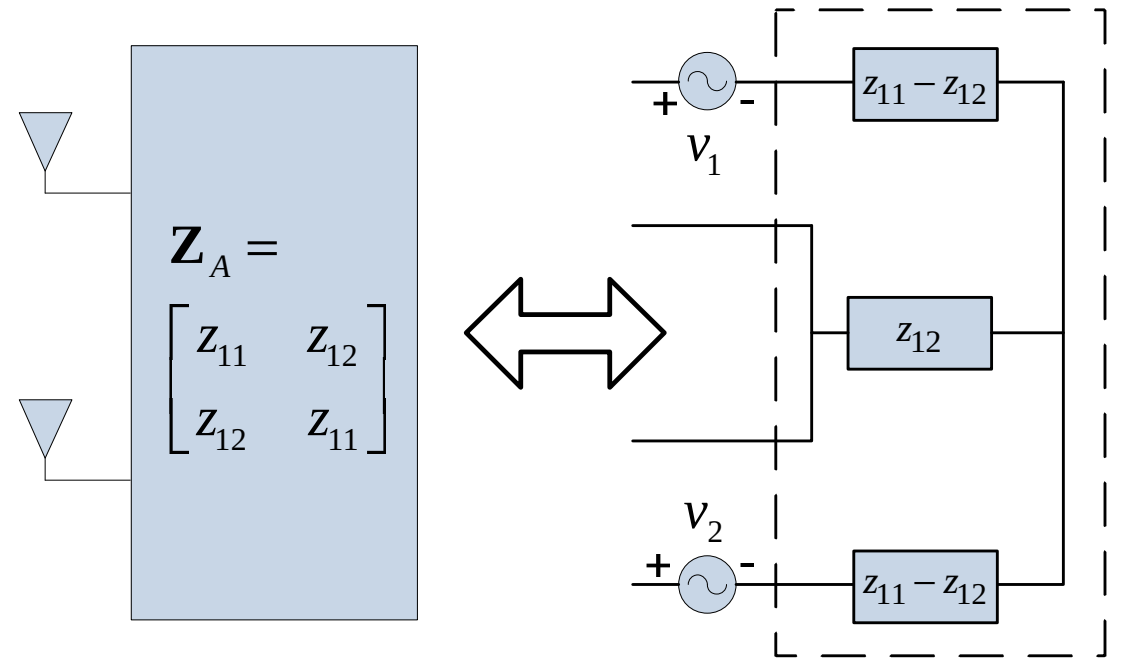

Fig. 2. Impedance matrix representation of a 2-element array

Although the antenna array is an $N$-port network, it can be appropriately extended to a $2N$-port network for mathematical convenience [4], such that

$$\mathbf{S}_A(j\omega) = \begin{bmatrix} \mathbf{S}_{22a}(j\omega) & \mathbf{S}_{21a}(j\omega) \\ \mathbf{S}_{21a}(j\omega) & \mathbf{S}_{22a}(j\omega) \end{bmatrix} ,$$

where the original $N$-port representation of the antenna array is represented by $\mathbf{S}_{22a}$ block, computed using

$$\mathbf{S}_{22a}(j\omega) = (\mathbf{Z}_A(j\omega) + \mathbf{I})^{-1}(\mathbf{Z}_A(j\omega) - \mathbf{I}) \ .$$

The other blocks must be evaluated based on the lossless ($\mathbf{S}_A^H\mathbf{S}_A = \mathbf{I}$) and reciprocal ($\mathbf{S}_A = \mathbf{S}_A^T$) properties of the antenna array. The symmetry of the system under consideration and that of the individual blocks constituting $\mathbf{S}_A$, further allows us to write (using EVD)

$$\mathbf{S}_A(j\omega) = \begin{bmatrix} \mathbf{Q} & \mathbf{0} \\ \mathbf{0} & \mathbf{Q} \end{bmatrix} \begin{bmatrix} \mathbf{\Lambda}_{22a}(j\omega) & \mathbf{\Lambda}_{21a}(j\omega) \\ \mathbf{\Lambda}_{21a}(j\omega) & \mathbf{\Lambda}_{22a}(j\omega) \end{bmatrix} \begin{bmatrix} \mathbf{Q}^H & \mathbf{0} \\ \mathbf{0} & \mathbf{Q}^H \end{bmatrix} ,$$

where

$$\mathbf{\Lambda}_{22a}(j\omega) \ = \ (\mathbf{\Lambda}_A(j\omega) + \mathbf{I})^{-1}(\mathbf{\Lambda}_A(j\omega) - \mathbf{I}) \ . \tag{7}$$

### B. Impedance Characterization

Next, we characterize the individual entries of antenna impedance matrix $\mathbf{Z}_A$ over a broad range of frequencies. Fano's broadband matching theory requires an impedance to be analytic over the entire $s$-plane. In order words, the resistive part must be an even function of $\omega$ and the reactive part an odd function of $\omega$. Since the eigen-impedances are a linear function of self and mutual-impedances, it suffices to consider an impedance model fit for the eigen-values of the antenna impedance matrix.

Although the resistive part of the eigen-impedances is, in general, frequency-dependent, for reasons outlined next, we assume that it is fairly constant over the frequency range of interest (say, $10\%$ relative bandwidth). First, the resistive part variation over frequency – as obtained from numerical EM code (NEC) simulations – will be shown to be much slower than the reactive part. Second, the system analysis is greatly simplified by such as assumption. The system transfer function, which is essentially governed by the antenna reflection coefficients (magnitude-squared), will be shown to closely approximate that of a series RLC model.[4] Thus, eigen-impedances assume the form $\lambda(j\omega) = R + jX(\omega)$, where $R$ and $X$ represent the real and imaginary parts, respectively.

[3] Although a unitary transformation for such matrices exists for each data-point over the frequency range, it is, in general, frequency-dependent.

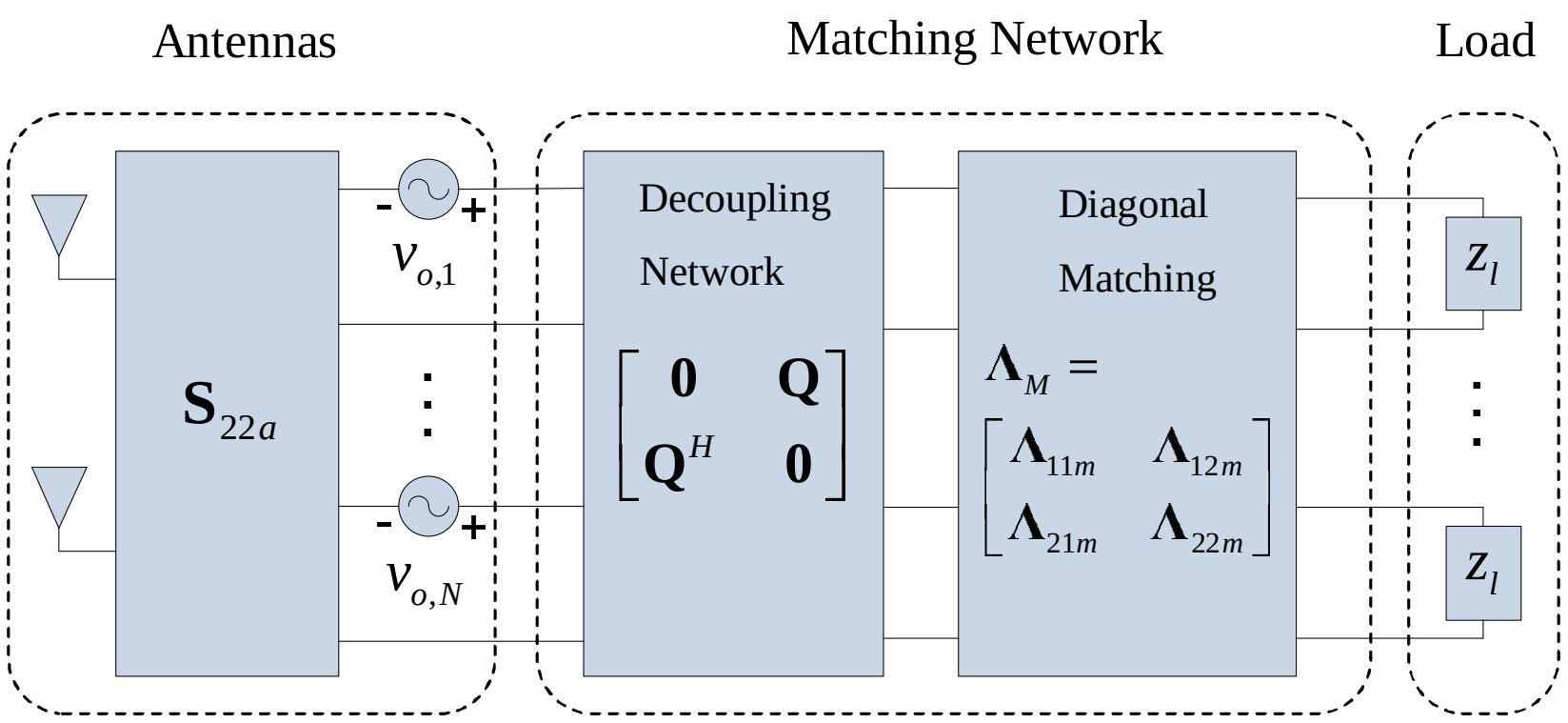

Fig. 3. Optimal matching implementation using decoupling networks

The eigen-impedances can essentially be regarded as *virtual antennas* comprising an uncoupled antenna array. In order to analyze their frequency responses, we must first characterize the eigen-impedances. Consider an array of two ideal and identical $\lambda/2$ resonant dipole antennas placed sufficiently far apart. Such a resonant antenna can be well modeled by a series RLC circuit within a $10\%$ relative bandwidth [16], [17]. For such an array, $\mathbf{Z}_A$ is diagonal. A resonant antenna with a series RLC equivalent, can alternatively be expressed in terms of its *resonant frequency* $\omega_0$ and *quality factor* $Q$, where:

$$\omega_0 = \frac{1}{\sqrt{LC}}\ ,\ Q = \frac{1}{R}\sqrt{\frac{L}{C}}\ .$$

As the spacing between the antennas decreases, EM interactions start to alter the spectral responses of the individual antennas. Numerical simulations using NEC suggest that the eigen-impedances also conform to that of a resonant antenna. Hence, we model them using a series RLC equivalent, or resonance parameters $(Q, \omega_0)$ as

$$\begin{aligned}\lambda_n(s) &= R_n + L_n s + \frac{1}{C_n s}\ ,\ R_n,\ L_n,\ C_n > 0\ ,\\ \lambda_n(j\omega) &= R_n\left[1 + jQ_n\left(\frac{\omega}{\omega_{0n}} - \frac{\omega_{0n}}{\omega}\right)\right]\ ,\end{aligned}$$

where $L_n = R_n Q_n/\omega_{0n}$ , $C_n = 1/Q_n R_n \omega_{0n}$. The corresponding reflection and transmission coefficients

$$\begin{aligned}\mathbf{\Lambda}_{11a}(s) &= \begin{bmatrix}\Gamma'_1(s) & 0\\ 0 & \Gamma'_2(s)\end{bmatrix}\ ,\\ \mathbf{\Lambda}_{21a}(s) &= \begin{bmatrix}T'_1(s) & 0\\ 0 & T'_2(s)\end{bmatrix}\ ,\end{aligned}$$

normalized to $R_n\ \Omega$ impedances are given by

$$\Gamma'_n(s) = \frac{1 + (s/\omega_{0n})^2}{1 + (s/\omega_{0n})^2 + 2(s/Q_n\omega_{0n})}\ , \tag{8}$$

$$T'_n(s) = \left(1 + \frac{Q_n}{2}\left(\frac{\omega_{0n}}{s} + \frac{s}{\omega_{0n}}\right)\right)^{-1}\ . \tag{9}$$

We refer to the frequency response of these virtual antennas, denoted by

$$|T'_n(f)|^2 = 1 - |\Gamma'_n(f)|^2 = \frac{4f^2}{4f^2 + Q_n^2(f^2 - f_{0n}^2)^2} \tag{10}$$

as *eigen-modes* of the coupled array.

### C. Broadband Matching Constraints

The broadband matching network design for this uncoupled system is rather straight-forward. The proposed matching network implementation is illustrated in Fig. 3. The purpose of matching is essentially shaping these spectral responses such that the overall system has a frequency-flat response over the bandwidth of interest. Ideally, one would expect to choose a matching network that ensures

$$|T_n(f)|^2 = 1 - |\Gamma_n(f)|^2,\ f\ \in\ B\ ,$$

is unity. Here, $T_n,\ \Gamma_n$ correspond to the transmission and reflection coefficient of the *cascade* of the antenna array and the matching network. However, Fano's broadband matching theory [18] reveals that there exist gain-bandwidth trade-offs for physically realizable matching networks, built using lumped passive elements. It imposes certain integral bounds – determined by the source and load impedances connecting the matching network – on the matching efficiency of the network.

For the series RLC model considered in our work, the set of broadband matching constraints are given by

$$(a) \int_B \log\frac{1}{|\Gamma_n(f)|^2}\ df = \frac{\omega_{0n}}{Q_n} - \sum_i z_{n,ri}\ , \tag{11a}$$

$$(b) \int_B f^{-2}\log\frac{1}{|\Gamma_n(f)|^2}\ df = \frac{4\pi^2}{\omega_{0n}Q_n} - \sum_i z_{n,ri}^{-1}\ , \tag{11b}$$

where $z_r$ represent additional zeros in $\Gamma_n$, that may sometimes be necessary to introduce in the right-half complex plane ($\mathrm{Re}(z_r) > 0$) in order to satisfy all of these constraints.[5]

[4] Without loss of generality, more rigorous impedance models can be employed to analyze such coupled MIMO systems using the approach outlined in this paper.

[5] $z_r$ must occur in conjugate pair if they are complex. The authors would like to point out a typo in [19], Eq. (14b); the correct bound is as stated above.

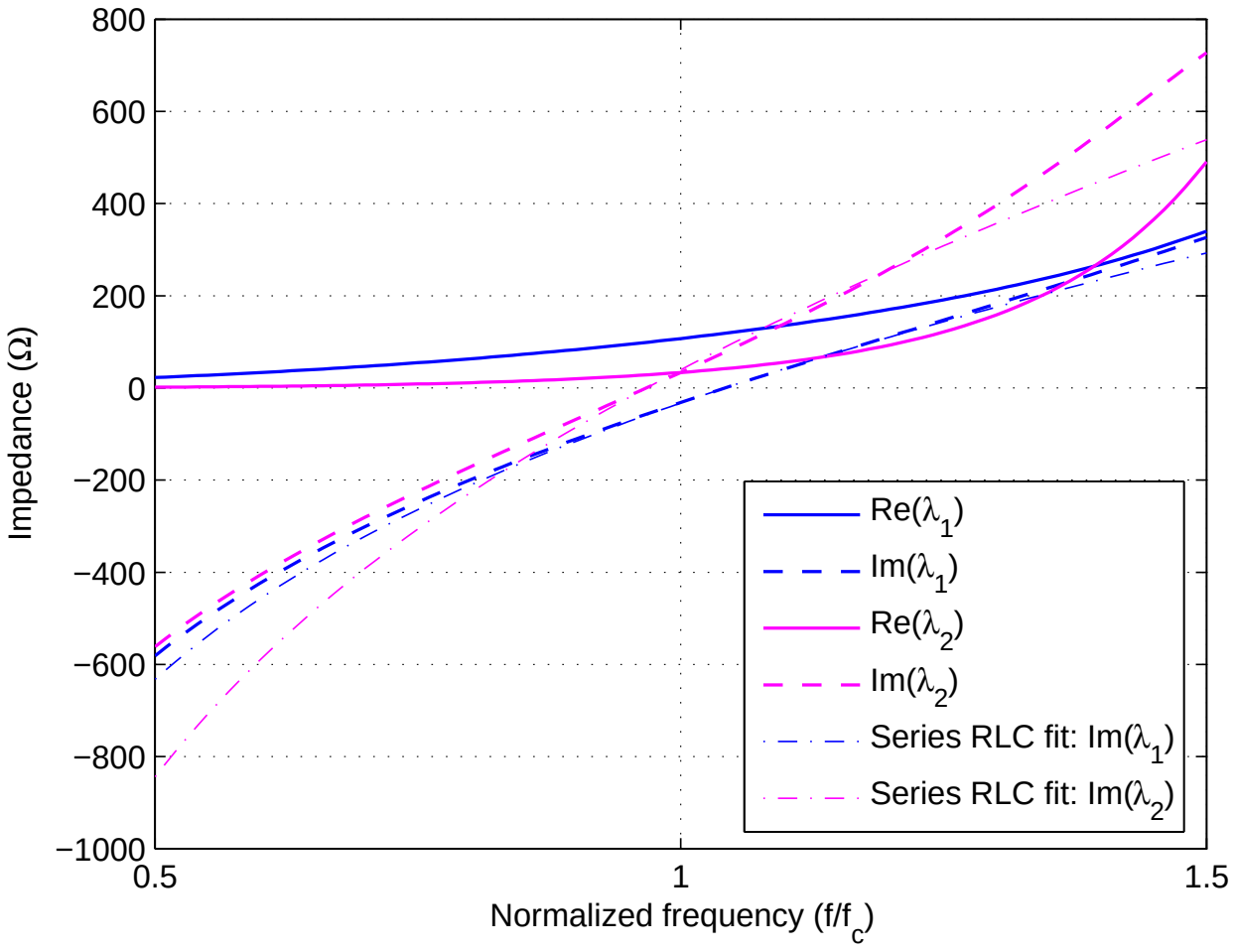

Fig. 4. Impedance: $N = 2$, $d = 0.25\lambda_c$

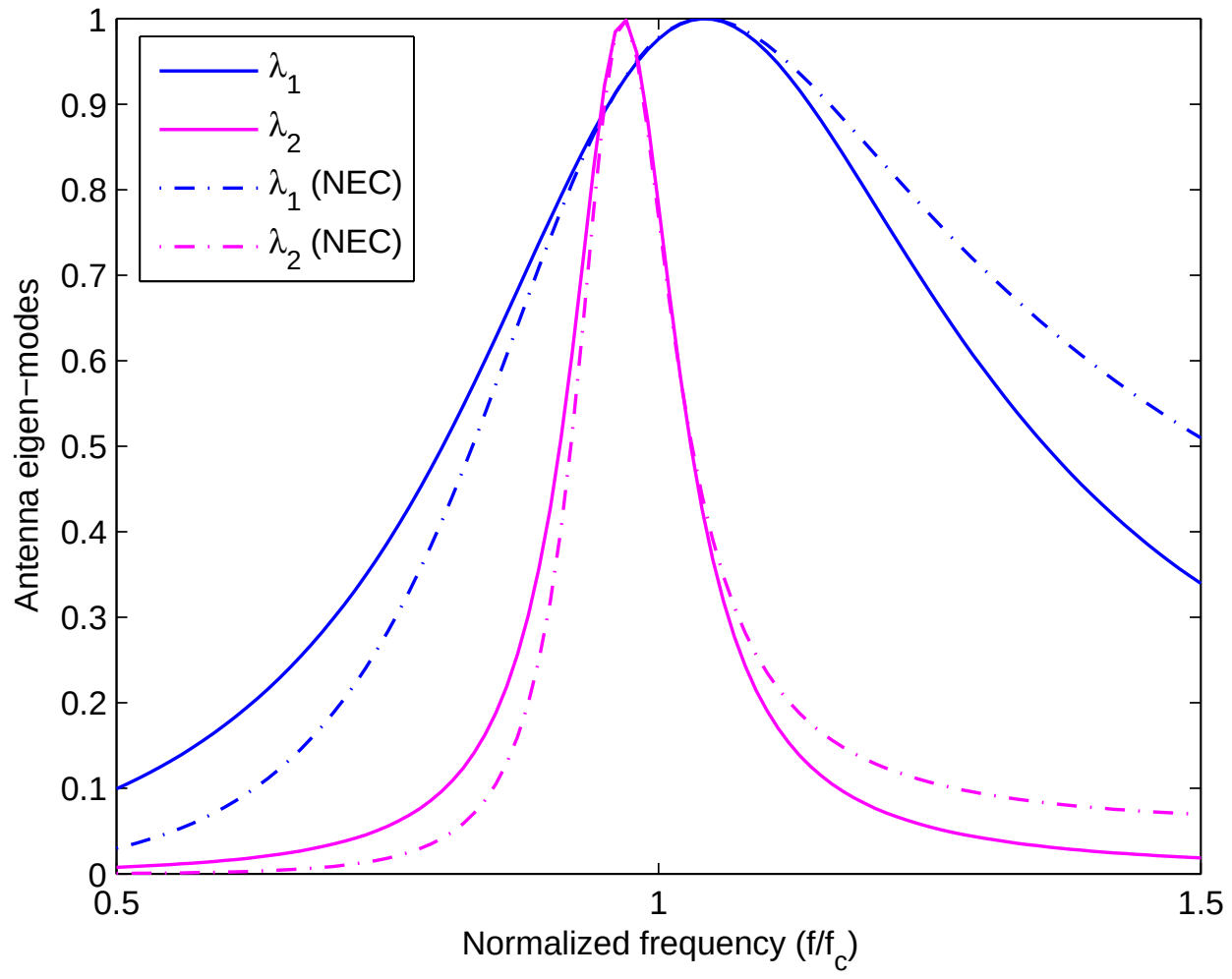

Fig. 5. Antenna eigen-modes: $N = 2$, $d = 0.25\lambda_c$

Observe how the bound is inversely proportional to the antenna $Q$. Clearly, an antenna with a broader frequency response (lower $Q$) offers better gain-bandwidth trade-offs.

## IV. Eigen-Mode Illustrations

Next we present numerical results for $N = 2,\ 3,\ 4$ element array of dipole antennas with length[6] $0.475\lambda_c$ and radius $10^{-3}\lambda_c$ for a variety of antenna spacings $d$.

### A. Impedance Parameters

Fig. 4 shows the broadband eigen-impedances calculated using NEC for $N = 2$ and $d = 0.25\lambda_c$. The resistive and reactive parts of the impedance are in general, non-linear functions of frequency. The figure also shows the imaginary part of eigen-impedances obtained by series RLC-fit which evidently, is in close agreement for bandwidths on the order of $10\%$. Although we consider a series RLC model for ease of analysis, the main results and ideas conveyed in this work can be applied to more rigorous models, as long as these models are analytic over the entire complex plane, in order to apply Fano's matching theory.

Fig. 5 shows the corresponding eigen-modes found by curve-fitting the resonance parameters $(Q, \omega_0)$ for the same setting. It clearly shows the impact coupling has on the bandwidth of the two virtual antennas. Mutual coupling is seen to decompose a two antenna coupled array with identical spatial modes into two spectrally non-identical eigen-modes – one *broadband* and the other *narrowband*. At much smaller spacings (e.g., $d = 0.1\lambda_c$), the narrower mode is essentially non-existent. The shrinking bandwidth of certain eigen-modes with increasing coupling is a manifestation of array super-directivity [20]. Table I summarizes this data for $N = 2$, $d = 0.25\lambda_c$.

Similar results have appeared in [12], where the common and difference modes (essentially, the eigen-modes) of a 2-antenna coupled array have been illustrated for $d = 0.1\lambda_c$, along with the orthogonal radiation patterns and a simulation test-bed to realize the system. [13] further extends the findings and implementation results to a 4-antenna circular array. It also proposes equivalent circuit models for each of the four eigen-modes using ladder LC networks, as opposed to the series RLC fit assumed in this paper.

[6]The length is chosen such that each antenna in isolation, has a relative resonance frequency of 1.

### B. Usable Bandwidths

Microwave/RF bandwidths are usually parameterized by *voltage standing-wave ratio* (VSWR): a metric based on matching efficiency, indicative of the range of voltage fluctuations in the standing wave formed due to reflections arising from an impedance-mismatch [21],

$$\mathrm{VSWR} = \frac{1 + |\Gamma(\omega)|}{1 - |\Gamma(\omega)|} .$$

The higher this ratio, the larger the mismatch and smaller the bandwidth. Note that $0 < |\Gamma(\omega)| < 1$, implies VSWR $\geq 1$. A convenient measure of the RF bandwidth is defined as the frequency range such that [21], [16], $1 \leq \text{VSWR} \leq 2$, i.e.,

$$|\Gamma(\omega)| \leq 1/3 \implies 1 - |\Gamma(\omega)|^2 \geq 8/9 \ .$$

TABLE I
Eigen-impedance parameters: $N = 2$, $d = 0.25\lambda_c$

| **Parameter** | **Value** |
|---|---|
| Antenna type | Dipole |
| Antenna length | $0.475\lambda_c$ |
| Antenna radius | $0.001\lambda_c$ |
| Quality factor ($Q_1$) | 3.75 |
| Resonant frequency ($f_{01}$) | $1.0425 f_c$ |
| $R_1$ | 118.76 $\Omega$ |
| $L_1 = Q_1 R_1/\omega_{01}$ | $67.99/f_c$ H |
| $C_1 = 1/Q_1 R_1 \omega_{01}$ | $342.78/f_c$ $\mu$F |
| Quality factor ($Q_2$) | 16 |
| Resonant frequency ($f_{02}$) | $0.9675 f_c$ |
| $R_2$ | 28.31 $\Omega$ |
| $L_2 = Q_2 R_2/\omega_{02}$ | $74.53/f_c$ H |
| $C_2 = 1/Q_2 R_2 \omega_{02}$ | $1.4/f_c$ mF |

Fig. 6 shows the VSWR and usable bandwidth for eigen-impedances calculated using NEC with termination $R_1$ and $R_2$, respectively.

### C. Discussion

These results shed substantial light on the impact coupling has in governing the RF bandwidth of compact arrays. It offers intuitively-coherent interpretations – as the spacing goes to zero, in limit, the narrowband mode must vanish, leaving behind a system that behaves like a single antenna system. The other key observation – that the two modes are centered at different resonant frequencies – explains that a multiport matching network designed with narrowband assumptions is sub-optimal and that physically realizable optimal networks ought to be optimized over the bandwidth. It also highlights the importance of antenna design that can exploit this phenomenon by trimming the antenna lengths and centering these modes so as to maximize the diversity gains.

Fig. 7–9 illustrate eigen-mode behaviors for different uniform circular array sizes $N = 2, 3, 4$, and antenna separations.[7] As expected, the impact of coupling at smaller spacings becomes profound as the array size grows. An analogy for the asymmetric behavior of eigen-modes can be drawn from quarter-wave transformers, a well known matching technique in microwave literature [22]. Each virtual antenna can be thought of as a transformer with a different length and impedance, thereby matching the other virtual antenna at a different frequency with a different bandwidth efficiency.

## V. Diversity-OFDM for Compact Arrays

Having analyzed how coupling impacts the RF bandwidth in the context of circular arrays, next, we evaluate the capacity of broadband diversity systems for variable antenna spacing. We choose OFDM as the broadband transmission scheme for our system. But first, we introduce the basic elements of multiport networks useful in modeling the coupled channel vector, followed by a system model.

### A. Broadband Coupled Array Model

To that end, consider the receive diversity system with $N$ antennas, as shown in Fig. 10 in its S-matrix network representation. It shows the cascade of two $2N$-port networks – $N_a$, representing a coupled lossless and reciprocal antenna array , and $N_m$ representing the lossless and reciprocal matching network – terminated into a bank of uncoupled load impedances $z_L$. The load here is indicative of low noise amplifiers and other downstream components of an RF chain, primarily, mixers and A/D converters. The EM field incident on the receive antenna array induces an open-circuit voltage across the antenna terminals, which acts as the *source excitation*, represented by the input wave vector $\mathbf{a}_1$ toward the left of the antenna array.

[7]In the context of a UCA, $d$ is defined as the separation between adjacent antennas.

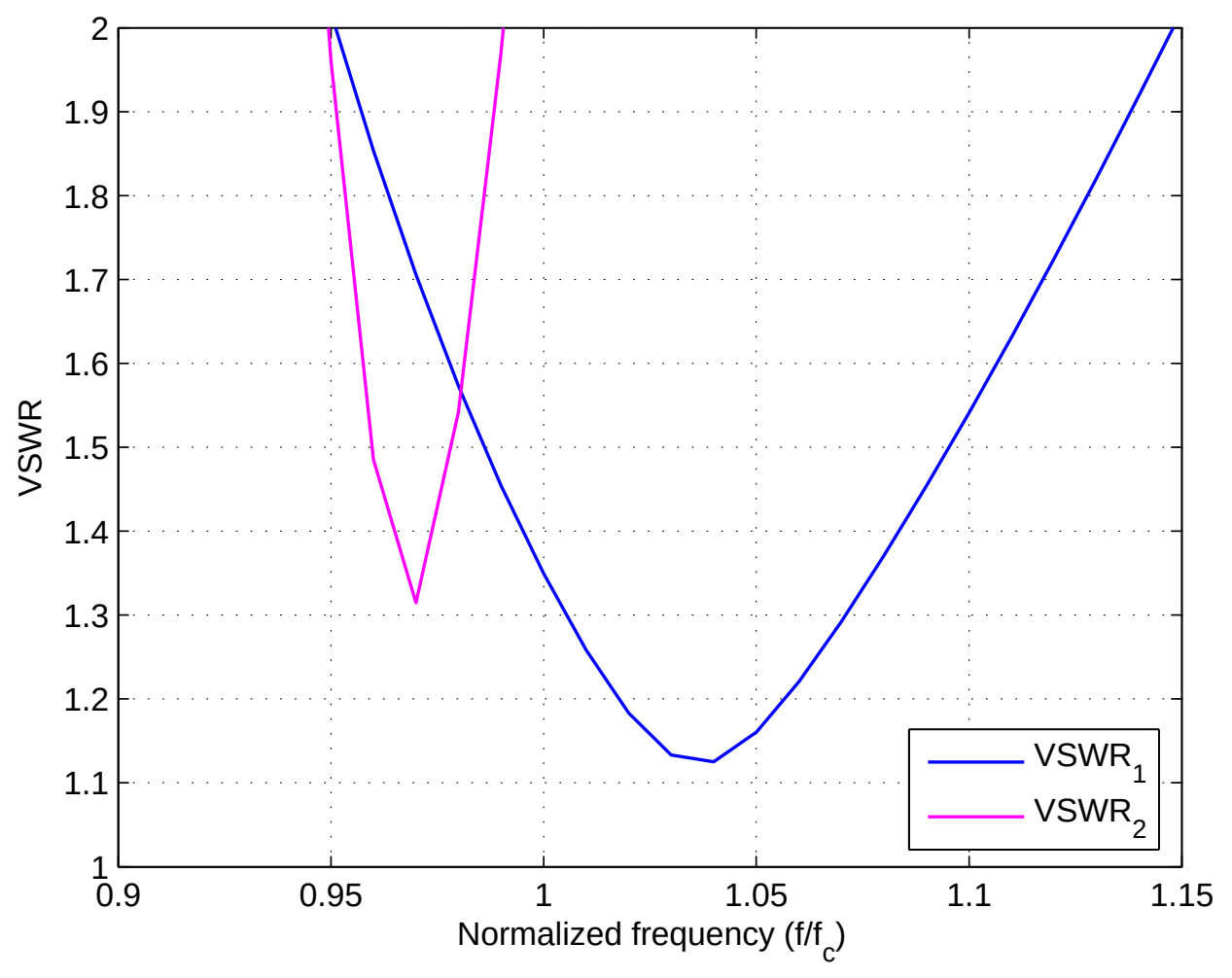

Fig. 6. VSWR of a 2-element array for $d = 0.25\lambda_c$

The $2N \times 2N$ S-matrices for the antenna array and matching network (in $N \times N$ block-matrix format) normalized with respect to 1 Ω reference impedances are given by

$$\mathbf{S}_A = \begin{bmatrix} \mathbf{S}_{11a} & \mathbf{S}_{12a} \\ \mathbf{S}_{21a} & \mathbf{S}_{22a} \end{bmatrix} ,$$

$$\mathbf{S}_M = \begin{bmatrix} \mathbf{S}_{11m} & \mathbf{S}_{12m} \\ \mathbf{S}_{21m} & \mathbf{S}_{22m} \end{bmatrix} ,$$

where we have omitted the frequency-dependence by suppressing $(s)$ for aesthetic reasons. We shall henceforth assume it implied, unless stated otherwise. The cascaded $2N$-port network has an S-matrix

$$\mathbf{S}_C = \mathbf{S}_A \otimes \mathbf{S}_M = \begin{bmatrix} \mathbf{S}_{11c} & \mathbf{S}_{12c} \\ \mathbf{S}_{21c} & \mathbf{S}_{22c} \end{bmatrix} ,$$

where, $\otimes$ represents the cascading operation and

$$\mathbf{S}_{11c} = \mathbf{S}_{11a} + \mathbf{S}_{12a}(\mathbf{I} - \mathbf{S}_{11m}\mathbf{S}_{22a})^{-1}\mathbf{S}_{11m}\mathbf{S}_{21a} \ , \quad (12a)$$

$$\mathbf{S}_{12c} = \mathbf{S}_{12a}(\mathbf{I} - \mathbf{S}_{11m}\mathbf{S}_{22a})^{-1}\mathbf{S}_{12m} \ , \quad (12b)$$

$$\mathbf{S}_{21c} = \mathbf{S}_{21m}(\mathbf{I} - \mathbf{S}_{22a}\mathbf{S}_{11m})^{-1}\mathbf{S}_{21a} \ , \quad (12c)$$

$$\mathbf{S}_{22c} = \mathbf{S}_{22m} + \mathbf{S}_{21m}(\mathbf{I} - \mathbf{S}_{22a}\mathbf{S}_{11m})^{-1}\mathbf{S}_{22a}\mathbf{S}_{12m} \ . \quad (12d)$$

The inward and outward traveling wave vectors are related by (1), except that, the input and output wave-vectors at the left are now vectors, denoted by $\mathbf{a}_1$ and $\mathbf{b}_1$, respectively. Those on the right side are denoted by $\mathbf{a}_2$ and $\mathbf{b}_2$, such that

$$\begin{bmatrix} \mathbf{b}_1 \\ \mathbf{b}_2 \end{bmatrix} = \begin{bmatrix} \mathbf{S}_{11c} & \mathbf{S}_{12c} \\ \mathbf{S}_{21c} & \mathbf{S}_{22c} \end{bmatrix} \begin{bmatrix} \mathbf{a}_1 \\ \mathbf{a}_2 \end{bmatrix} . \quad (13)$$

It can be easily shown that optimal matching combined with the decoupling network (spatial-DFT) can be realized by

$$\mathbf{S}_M = \begin{bmatrix} \mathbf{Q} & \mathbf{0} \\ \mathbf{0} & \mathbf{I} \end{bmatrix} \begin{bmatrix} \mathbf{\Lambda}_{11m} & \mathbf{\Lambda}_{12m} \\ \mathbf{\Lambda}_{21m} & \mathbf{\Lambda}_{22m} \end{bmatrix} \begin{bmatrix} \mathbf{Q}^H & \mathbf{0} \\ \mathbf{0} & \mathbf{I} \end{bmatrix} ,$$

such that from (12), the overall cascaded network can be rewritten in block-matrix format as

$$\mathbf{S}_C = \begin{bmatrix} \mathbf{Q} & \mathbf{0} \\ \mathbf{0} & \mathbf{Q} \end{bmatrix} \begin{bmatrix} \mathbf{\Gamma} & \mathbf{T} \\ \mathbf{T} & \mathbf{\Gamma} \end{bmatrix} \begin{bmatrix} \mathbf{Q}^H & \mathbf{0} \\ \mathbf{0} & \mathbf{Q}^H \end{bmatrix} ,$$

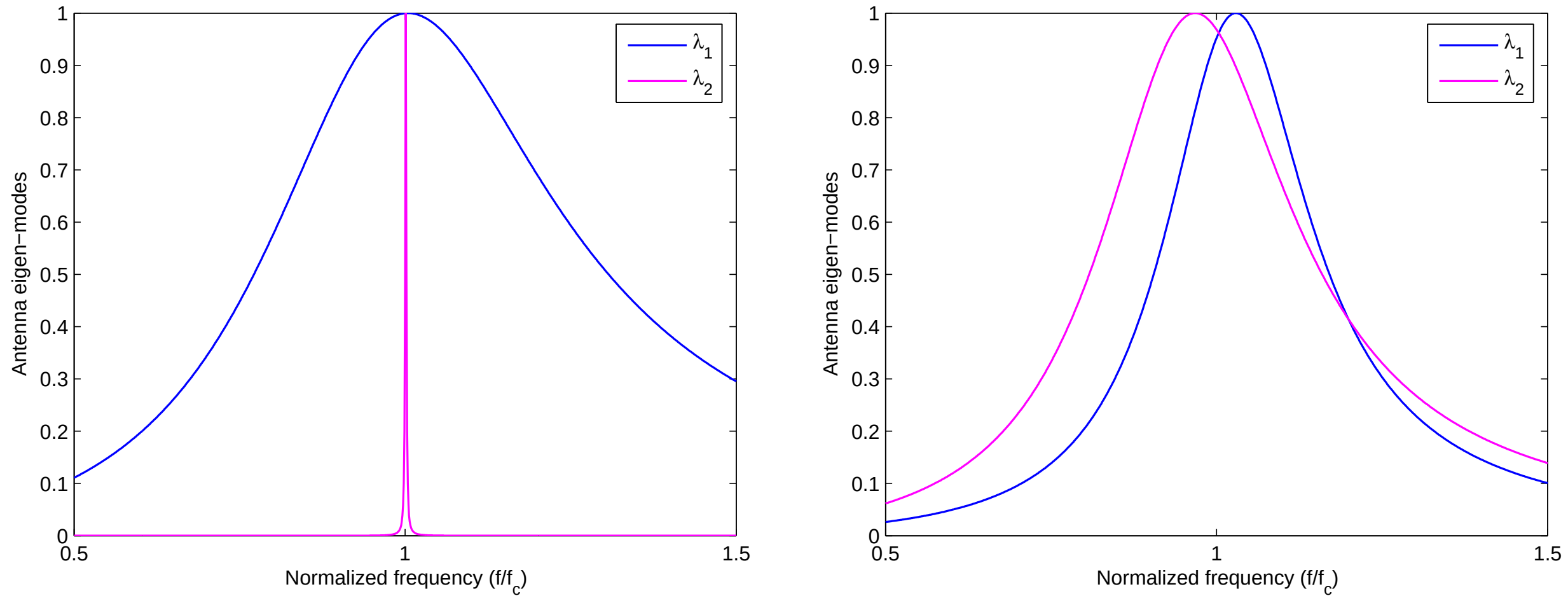

(a) $d = 0.1\lambda_c$ (b) $d = 0.5\lambda_c$

Fig. 7. Antenna eigen-modes: $N = 2$

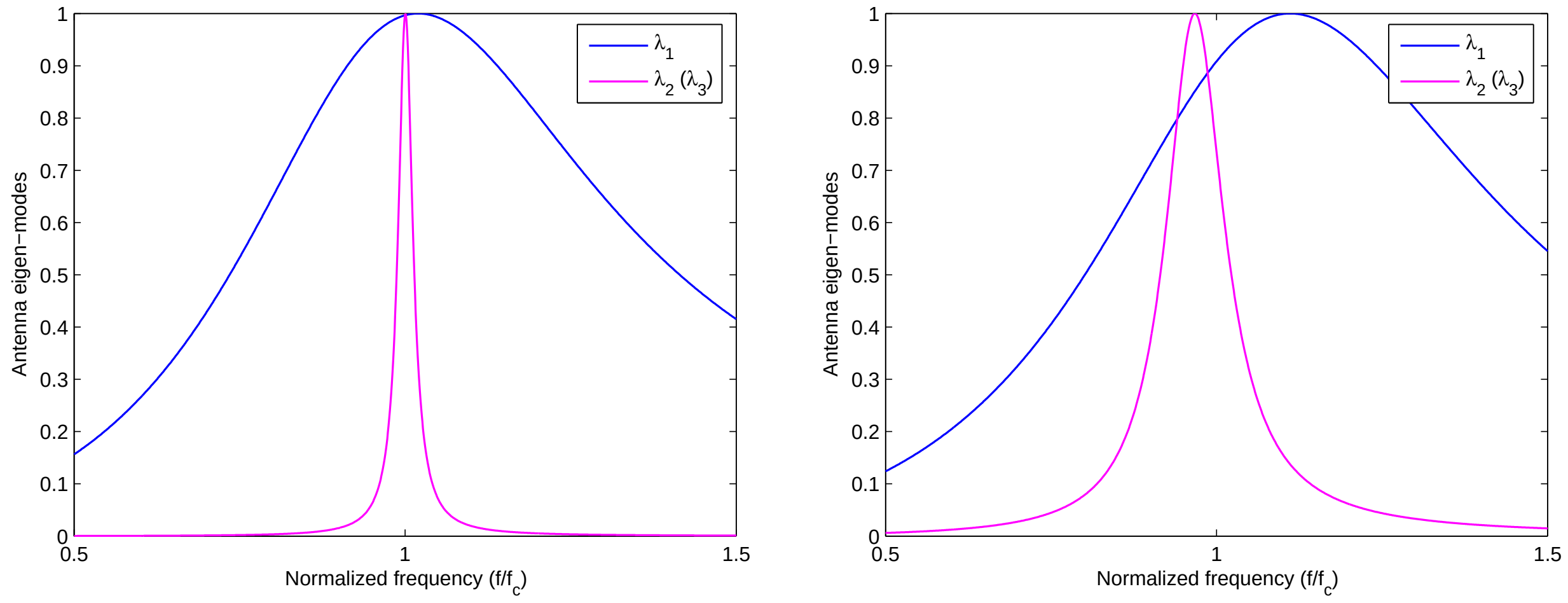

(a) $d = 0.1\lambda_c$ (b) $d = 0.25\lambda_c$

Fig. 8. Antenna eigen-modes: $N = 3$

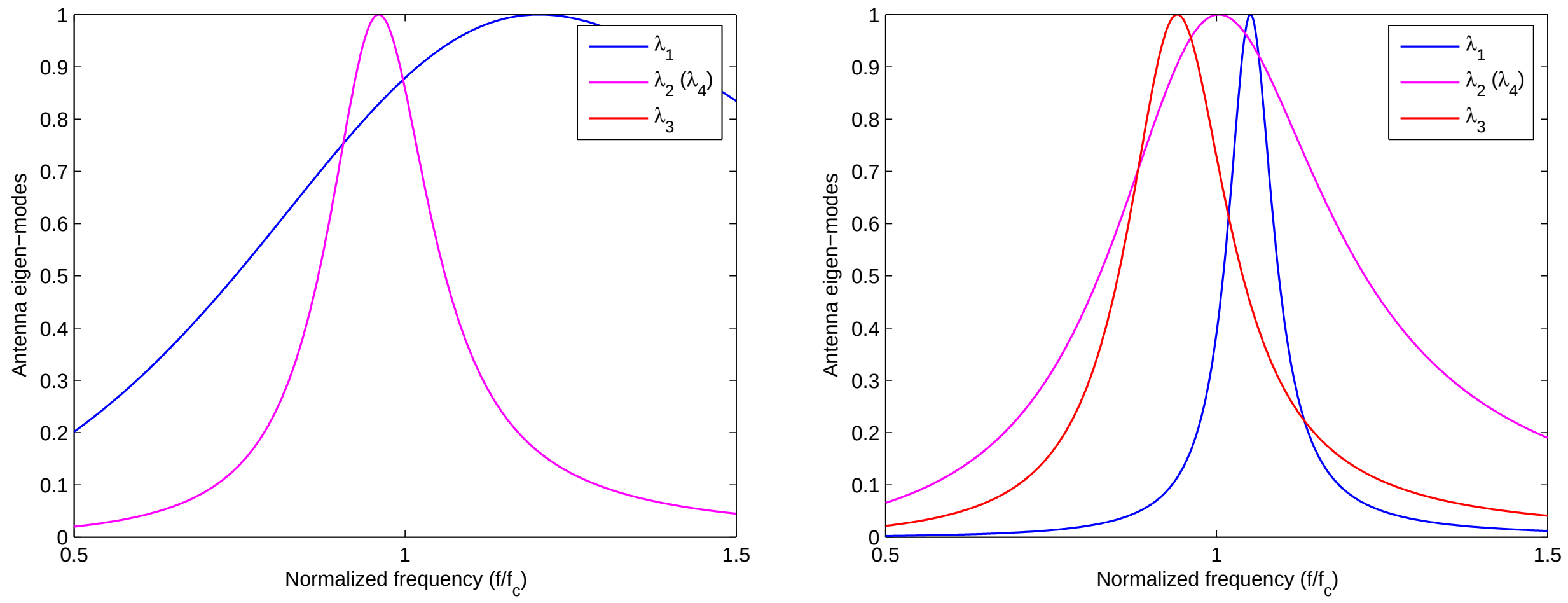

(a) $d = 0.25\lambda_c$ (b) $d = 0.5\lambda_c$

Fig. 9. Antenna eigen-modes: $N = 4$

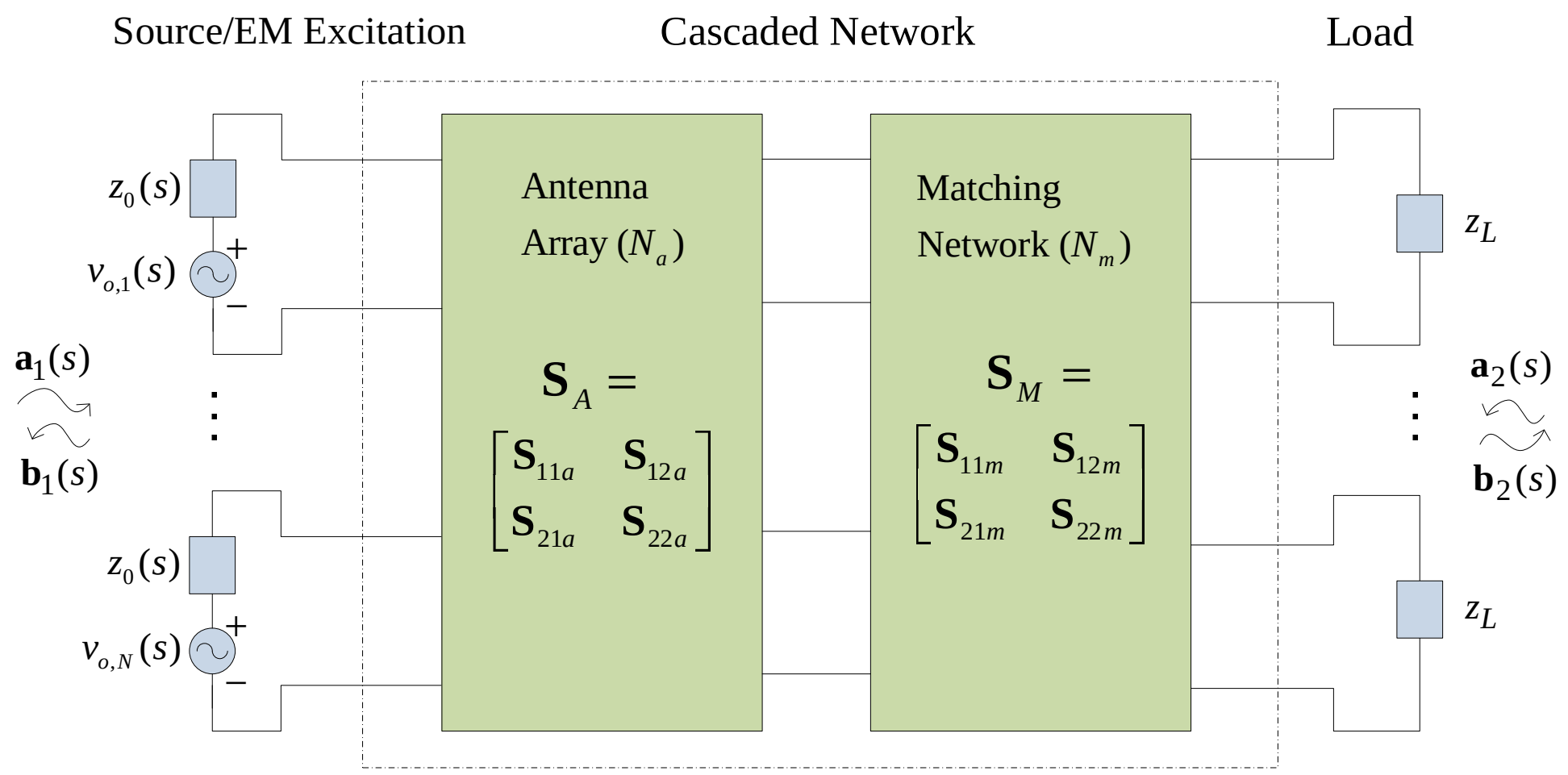

Fig. 10. S-matrix representation of RF front-end

where,

$$\mathbf{\Gamma} \;\; = \;\; \mathbf{\Lambda}_{11a} + \mathbf{\Lambda}_{12a}(\mathbf{I} - \mathbf{\Lambda}_{11m}\mathbf{\Lambda}_{22a})^{-1}\mathbf{\Lambda}_{11m}\mathbf{\Lambda}_{21a} \ .$$

From a broadband matching perspective, it suffices to analyze the constraints on the eigen-values of $\mathbf{S}_{11c}$ given above. This is once again equivalent to saying that the problem of matching the coupled identical antenna array $\mathbf{Z}_A$ has been replaced by that of a uncoupled non-identical *virtual antenna array* $\mathbf{\Lambda}_A$. The entries of

$$\mathbf{\Gamma}(s) \;\; = \;\; \begin{bmatrix} \Gamma_1(s) & 0 \\ 0 & \Gamma_2(s) \end{bmatrix} \ ,$$

represent the reflection coefficients at the output of the matching network.

### B. Signal Model

We begin by presenting the signal model of a traditional Diversity-OFDM system with 1 transmit and $N$ receive antennas, which we shall later extend to incorporate coupling. It is assumed that the separation between the antennas is such that coupling between them is negligible.

It is well known that the use of orthogonal sub-carriers in OFDM with a cyclic prefix converts a frequency-selective MIMO channel into a set of parallel frequency-flat MIMO channels [23]. A Diversity-OFDM system with $K$ sub-carriers (spanning bandwidth $B$) modulated by symbols $s_k$ where $k$ represents the $k$-th sub-carrier, is well modeled by

$$\mathbf{r}_k = \mathbf{h}_k s_k + \mathbf{n}_k \ , \ k = 1, \ldots, K. \tag{14}$$

Here, $\mathbf{r}_k$ is the $N \times 1$ received vector symbol on the $k$-th sub-carrier and $\mathbf{n}_k$ is the $N \times 1$ additive white Gaussian noise (AWGN) vector at the receiver with zero mean and covariance[8] $\mathbf{R}_{\mathbf{n}_k} = \mathbb{E}[\mathbf{n}_k \mathbf{n}_k^H]$, denoted by $\mathbf{n}_k \sim \mathcal{CN}(\mathbf{0}, \mathbf{R}_{\mathbf{n}_k})$. The channel vector for the $k$-th sub-carrier is given by

$$\mathbf{h}_k \;\; = \;\; \begin{bmatrix} h_1[k] \\ \vdots \\ h_N[k] \end{bmatrix} \ , \ k = 1, \ldots, K.$$

[8] $\mathbb{E}[\cdot]$ represents the expectation operator.

The transmit and receive spatial fading-correlation is modeled using the Kronecker model [24] such that the $l$-th tap time-domain channel vector (obtained via inverse-Fourier transform) can be expressed as

$$\widetilde{\mathbf{h}}_l = \mathbf{R}_{\mathbf{h}}^{1/2} \widetilde{\mathbf{h}}_{wl} \tag{15}$$

where $\widetilde{\mathbf{h}}_{wl}$ represents the $1 \times N$ white channel vector having i.i.d. complex Gaussian entries with zero mean and unit variance $\widetilde{\mathbf{h}}_{wl} \sim \mathcal{CN}(\mathbf{0}, \mathbf{I})$, and $\mathbf{R}_{\mathbf{h}} = (1/N)\mathbb{E}[\widetilde{\mathbf{h}}_l \widetilde{\mathbf{h}}_l^H]$ denotes the receiver correlation.

Owing to the orthogonal decomposition of the frequency selective channel, the cumulative Diversity-OFDM system can be represented in matrix notation by

$$\boldsymbol{r} = \boldsymbol{\mathcal{H}} \boldsymbol{s} + \boldsymbol{n} \ , \tag{16}$$

where, $\boldsymbol{\mathcal{H}}$ is a $KN \times K$ block diagonal matrix given by

$$\boldsymbol{\mathcal{H}} = \begin{bmatrix} \mathbf{h}_1 & & \\ & \ddots & \\ & & \mathbf{h}_K \end{bmatrix} .$$

The $KN \times 1$ received vector $\boldsymbol{r}$, $K \times 1$ transmit vector $\boldsymbol{s}$ and $KN \times 1$ noise vector $\boldsymbol{n}$ are given by

$$\boldsymbol{r} = \begin{bmatrix} \mathbf{r}_1 \\ \vdots \\ \mathbf{r}_K \end{bmatrix} \ , \ \boldsymbol{s} = \begin{bmatrix} s_1 \\ \vdots \\ s_K \end{bmatrix} \ , \ \boldsymbol{n} = \begin{bmatrix} \mathbf{n}_1 \\ \vdots \\ \mathbf{n}_K \end{bmatrix} .$$

The Shannon capacity for such a system (in nats/s/Hz) is given by [25]

$$C = \frac{1}{K} \max_{\boldsymbol{\mathcal{R}_s}} \left\{ \log\det\left( \boldsymbol{I} + \boldsymbol{\mathcal{R}}_{\boldsymbol{n}}^{-1} \boldsymbol{\mathcal{H}} \boldsymbol{\mathcal{R}}_{\boldsymbol{s}} \boldsymbol{\mathcal{H}}^H \right) \right\} \tag{17}$$

where, $\boldsymbol{I}$ is an $NK \times NK$ identity matrix, and $\boldsymbol{\mathcal{R}}_{\boldsymbol{s}}$ and $\boldsymbol{\mathcal{R}}_{\boldsymbol{n}}$ are the transmit-signal and noise covariances:

$$\boldsymbol{\mathcal{R}}_{\boldsymbol{s}} = \mathbb{E}[\boldsymbol{s}\boldsymbol{s}^H] \ , \ \boldsymbol{\mathcal{R}}_{\boldsymbol{n}} = \mathbb{E}[\boldsymbol{n}\boldsymbol{n}^H] \ . \tag{18}$$

There have been numerous extensions to the above model in order to account for some of the channel non-idealities, such as fading correlation at the transmitter and/or the receiver, either due to smaller antenna separation or non-richness of the multipath fading environment. The broadband antenna behavior, however, has mostly been assumed ideal. In the next section, we introduce a channel model that incorporates antenna coupling into the signal-model outlined in (14).

### C. Coupled Signal Model

To model the impact of coupling, $\mathbf{h}_k$ must be modified to include the *transmission matrix* or *transmissivity* of the cascaded network (of antenna array and matching network) modeled by $\mathbf{S}_{21c}(f_k)$. We skip the tedious network analysis and present the effective channel vector at the $k$-th sub-carrier directly[9]:

$$\begin{aligned}\mathbf{h}'_k &= \mathbf{S}_{21m,k}(\mathbf{I}-\mathbf{S}_{22a,k}\mathbf{S}_{11m,k})^{-1}\mathbf{S}_{21a,k}\mathbf{h}_k \\ &= \mathbf{S}_{21c,k}\mathbf{h}_k \ . \end{aligned} \tag{19}$$

In the presence of mutual coupling, the signal model can thus be expressed for the $k$-th sub-carrier as

$$\mathbf{r}_k = \mathbf{S}_k\mathbf{h}_k s_k + \mathbf{n}_k \ , \tag{20}$$

where $\mathbf{S}_k \triangleq \mathbf{S}_{21c,k} = \mathbf{S}_{21c}(f_k)$.

### D. Receiver Noise Model

The additive noise $\mathbf{n}_k$ is usually modeled as a combination of noise from various sources in the RF chain [7]. In general, the noise sources can be categorized into three types: (a) *sky noise* or antenna noise, consisting of thermal radiation, cosmic background, and interference from other devices, (b) *amplifier noise*, and (c) *downstream noise*, consisting of noise from the rest of the RF chain components.

Alternatively, we classify the noise as antenna noise, and load noise – a combination of amplifier and downstream noise [27]. Furthermore, the load noise in general can be considered to be a combination of *forward traveling* noise $\mathbf{n}_f$, and *reverse traveling* noise $\mathbf{n}_r$. Thus, the total noise at $k$-th sub-carrier referenced to the load, is given by

$$\mathbf{n}_k = \mathbf{S}_{21c,k}\mathbf{n}_{s,k} + \mathbf{n}_{f,k} + \mathbf{S}_{22c,k}\mathbf{n}_{r,k} \ . \tag{21}$$

The sky noise and load noise can be well modeled as statistically independent, zero-mean, circularly symmetric, complex Gaussian (ZMCSCG) and spectrally white:

$$\begin{aligned}&\mathbf{n}_s \sim \mathcal{CN}(\mathbf{0}, 4k_B T_A B \mathbf{R}_A) \ , \ \mathbf{n}_f \sim \mathcal{CN}(\mathbf{0}, 4k_B T_f B\mathbf{I}) \ , \\ &\mathbf{n}_r \sim \mathcal{CN}(\mathbf{0}, 4k_B T_r B\mathbf{I}) \ ,\end{aligned}$$

a reasonable assumption for bandwidths less than $10\%$. Here, $T_A$ denotes the antenna temperature in Kelvin, while $T_f$ and $T_r$ are the *effective noise temperatures* which can be computed from the amplifier noise parameters (cf. [28, Chap. 1]).

The reverse and forward traveling noise waves are in general, correlated to an extent determined by exact amplifier models (cf. [28, Chap. 1]). and the noise covariance by

$$\begin{aligned}\mathbf{R}_{\mathbf{n}_k} &= 4k_B B\big[T_A\mathbf{S}_k\mathbf{R}_A\mathbf{S}_k^H + T_f\mathbf{I} + \dots \\ &\qquad \dots + T_r(\mathbf{I}-\mathbf{S}_k\mathbf{S}_k^H) - 2\,\mathrm{Re}\,(T_c^*\mathbf{S}_k)\,\big]\end{aligned}$$

where, $k_B$ is Boltzmann constant, $B$ is the system bandwidth, $\mathbf{R}_A = \mathrm{Re}(\mathbf{Z}_A)$ is frequency independent (by assumption), and $T_c$ represents the correlation between forward and reverse traveling noise. Due to the limited scope of this paper, we assume $T_c$ is negligible compared to $T_f$ and $T_r$, and restrict the impact of various noise sources by varying $T_A$, $T_f$ and $T_r$ relative to the standard temperature $T_0$.

Observe that $\mathbf{S}_k$ and $\mathbf{R}_A$ admit the familiar EVD

$$\mathbf{S}_k = \mathbf{Q}\mathbf{T}_k\mathbf{Q}^H \ , \ \mathbf{R}_A = \mathbf{Q}\,\mathrm{Re}(\mathbf{\Lambda}_A)\mathbf{Q}^H \ . \tag{22}$$

This allows us to diagonalize the noise covariance by $\mathbf{Q}$, i.e.,

$$\mathbf{R}_{\mathbf{n}_k} = \mathbf{Q}\mathbf{\Sigma}_{\mathbf{n}_k}\mathbf{Q}^H \ ,$$

such that,

$$\begin{aligned}\mathbf{\Sigma}_{\mathbf{n}_k} &= 4k_B B\big[(T_A - T_r)\,\mathrm{Re}(\mathbf{\Lambda}_A)(\mathbf{I}-\mathbf{\Gamma}_k\mathbf{\Gamma}_k^H) + \dots \\ &\qquad \dots + (T_f + T_r)\mathbf{I}\big] \ ,\end{aligned} \tag{23}$$

where we have used the lossless property of the network, i.e., $\mathbf{T}_k\mathbf{T}_k^H = \mathbf{I} - \mathbf{\Gamma}_k\mathbf{\Gamma}_k^H$. We normalize the noise covariance such that for i.i.d. case, $\mathbf{\Sigma}_{\mathbf{n}_k} = N_0\mathbf{I}$ :

$$N_0 = 4k_B B\left(T_A\,\mathrm{Re}(z_A)(1-|\Gamma_{iid}|^2) + T_f + T_r|\Gamma_{iid}|^2\right) \ .$$

### E. Capacity

The cumulative Diversity-OFDM system in the presence of mutual coupling can be written in matrix format (similar to (16)), as

$$\boldsymbol{r} = \boldsymbol{\mathcal{S}}\boldsymbol{\mathcal{H}}\boldsymbol{s} + \boldsymbol{n} \tag{24}$$

where $\boldsymbol{\mathcal{S}}$ is the $KN \times KN$ block-diagonal matrix given by

$$\boldsymbol{\mathcal{S}} = \begin{bmatrix} \mathbf{S}_1 & & \\ & \ddots & \\ & & \mathbf{S}_K \end{bmatrix} .$$

The Shannon capacity that incorporates the impact of mutual coupling can thus be represented by

$$C = \frac{1}{K}\max_{\boldsymbol{\mathcal{R}_s},\boldsymbol{\mathcal{S}}}\left\{\log\det\left(\boldsymbol{I} + \frac{1}{N}\boldsymbol{\mathcal{R}}_{\boldsymbol{n}}^{-1}\boldsymbol{\mathcal{S}}\boldsymbol{\mathcal{H}}\boldsymbol{\mathcal{R}_s}\boldsymbol{\mathcal{H}}^H\boldsymbol{\mathcal{S}}^H\right)\right\}$$

where the optimization space now also includes the matching network design. A recent study has analyzed information-theoretic limits of such a system in the presence of channel state information (CSI) that jointly optimizes transmit power allocation and receiver broadband matching [29] – such that the optimal solution follows a mutual space-frequency water-pouring characteristic.

In this work, we address a receiver design that operates fairly well over the entire signal bandwidth and is independent of the power allocation and channel fading conditions. By employing appropriate decoupling networks, the capacity subject to uniform power allocation across $K$ sub-carriers ($\boldsymbol{\mathcal{R}_s} = \mathcal{E}_s\boldsymbol{I}$), can be expressed as

$$\begin{aligned}C &= \frac{1}{K}\max_{\boldsymbol{\mathcal{S}}}\log\det\left(\boldsymbol{I} + \frac{\mathcal{E}_s}{N_0}\ \boldsymbol{\mathcal{R}}_{\boldsymbol{n}}^{-1}\boldsymbol{\mathcal{S}}\boldsymbol{\mathcal{H}}\boldsymbol{\mathcal{H}}^H\boldsymbol{\mathcal{S}}^H\right) \\ &= \frac{1}{K}\max_{\boldsymbol{\mathcal{S}}}\log\det\left(\boldsymbol{I} + \frac{\mathcal{E}_s}{N_0}\ \boldsymbol{\mathcal{H}}^H\boldsymbol{\mathcal{Q}}\boldsymbol{\Sigma}_{\boldsymbol{n}}^{-1}\boldsymbol{\mathcal{T}}\boldsymbol{\mathcal{T}}^H\boldsymbol{\mathcal{Q}}^H\boldsymbol{\mathcal{H}}\right)\end{aligned}$$

[9]Interested reader is referred to [26].

where we have expressed $\boldsymbol{\mathcal{S}}$ and $\boldsymbol{\mathcal{R}_n}$ as:

$$\boldsymbol{\mathcal{S}} = \boldsymbol{\mathcal{Q}}\boldsymbol{T}\boldsymbol{\mathcal{Q}}^H \ , \ \boldsymbol{\mathcal{R}_n} = \boldsymbol{\mathcal{Q}}\boldsymbol{\Sigma_n}\boldsymbol{\mathcal{Q}}^H \ ,$$

using $KN \times KN$ block-diagonal matrices

$$\boldsymbol{\mathcal{Q}} = \begin{bmatrix} \mathbf{Q} & & \\ & \ddots & \\ & & \mathbf{Q} \end{bmatrix} , \ \boldsymbol{T} = \begin{bmatrix} \mathbf{T}_1 & & \\ & \ddots & \\ & & \mathbf{T}_K \end{bmatrix} ,$$

$$\boldsymbol{\Sigma_n} = \begin{bmatrix} \boldsymbol{\Sigma}_{\mathbf{n}_1} & & \\ & \ddots & \\ & & \boldsymbol{\Sigma}_{\mathbf{n}_K} \end{bmatrix} .$$

The capacity can thus be simplified to

$$C = \frac{1}{K}\sum_{k=1}^{K} \max_{\boldsymbol{\Gamma}_k} \log\left(1 + \frac{\mathcal{E}_s}{N_0}\, \widehat{\mathbf{h}}_k^H \boldsymbol{\Sigma}_{\mathbf{n}_k}^{-1}(\mathbf{I} - \boldsymbol{\Gamma}_k\boldsymbol{\Gamma}_k^H)\widehat{\mathbf{h}}_k\right)$$

where, $\widehat{\mathbf{h}}_k = \mathbf{Q}^H\mathbf{h}_k$ represents the effective fading path-gains.

A simple and practical solution to the problem at hand is a *box-car* matching characteristic at the receiver, defined as:

$$\Gamma(f) = \begin{cases} \Gamma_0, & f \in B \\ 1, & \text{elsewhere} \end{cases}$$

such that

$$C = \frac{1}{K}\sum_{k=1}^{K} \log\left(1 + \frac{\mathcal{E}_s}{N_0}\, \widehat{\mathbf{h}}_k^H \boldsymbol{\Sigma}_{\mathbf{n}_k}^{-1}(\mathbf{I} - \boldsymbol{\Gamma}_k\boldsymbol{\Gamma}_k^H)\widehat{\mathbf{h}}_k\right) .$$

## VI. Results

Monte-Carlo simulations are carried out for $100,000$ channel realizations. We assume a quasi-static (or block) fading channel, i.e., the channel remains constant during each OFDM symbol. The fading paths gains are modeled as i.i.d. complex Gaussian entries with zero mean and unit variance, $\widetilde{\mathbf{h}}_{wl} \sim \mathcal{CN}(\mathbf{0}, \mathbf{I})$. For an $N$-antenna uniform circular array, the incident electric field is modeled in NEC as a superposition of $K' = 32$ vertically polarized plane waves with i.i.d. phases uniformly distributed on $[0, 2\pi)$. The angles-of-arrival (AOA) of the plane waves, $\phi_0, \ldots, \phi_{K'-1}$, are uniformly spaced in azimuth from $0$ to $2\pi$. Under these conditions, the open-circuit fading path gains for $m$-th and $n$-th antenna separated by $d_{mn}$ are approximately Gaussian with correlation matrix (cf. (15))

$$[\mathbf{R_h}]_{nm} = \sum_{k=0}^{K'-1} g_n(\phi_k) g_m^*(\phi_k) e^{j2\pi d_{nm}\cos(\phi_k)/\lambda} \ ,$$

where $g_n(\phi)$ is the open-circuit voltage induced in the $n$th antenna by a zero-phase plane wave with AOA $\phi$, normalized so that $\sum_k |g_n(\phi_k)|^2 = 1$ for an isolated dipole.

For each virtual antenna $\boldsymbol{\Gamma}_k$, we consider a box-car matching profile over a relative bandwidth $W = B/f_c$, such that the matching constraints (11) manifest themselves as

$$(a) \ W \log\frac{1}{|\Gamma_0|^2} = \frac{2\pi}{Q} - \frac{1}{f_0}\sum_i z_{ri} \ , \quad \text{(25a)}$$

$$(b) \ \frac{W}{(1 - W^2/4)} \log\frac{1}{|\Gamma_0|^2} = \frac{2\pi}{Q} - \frac{1}{f_0}\sum_i z_{ri}^{-1} \ . \quad \text{(25b)}$$

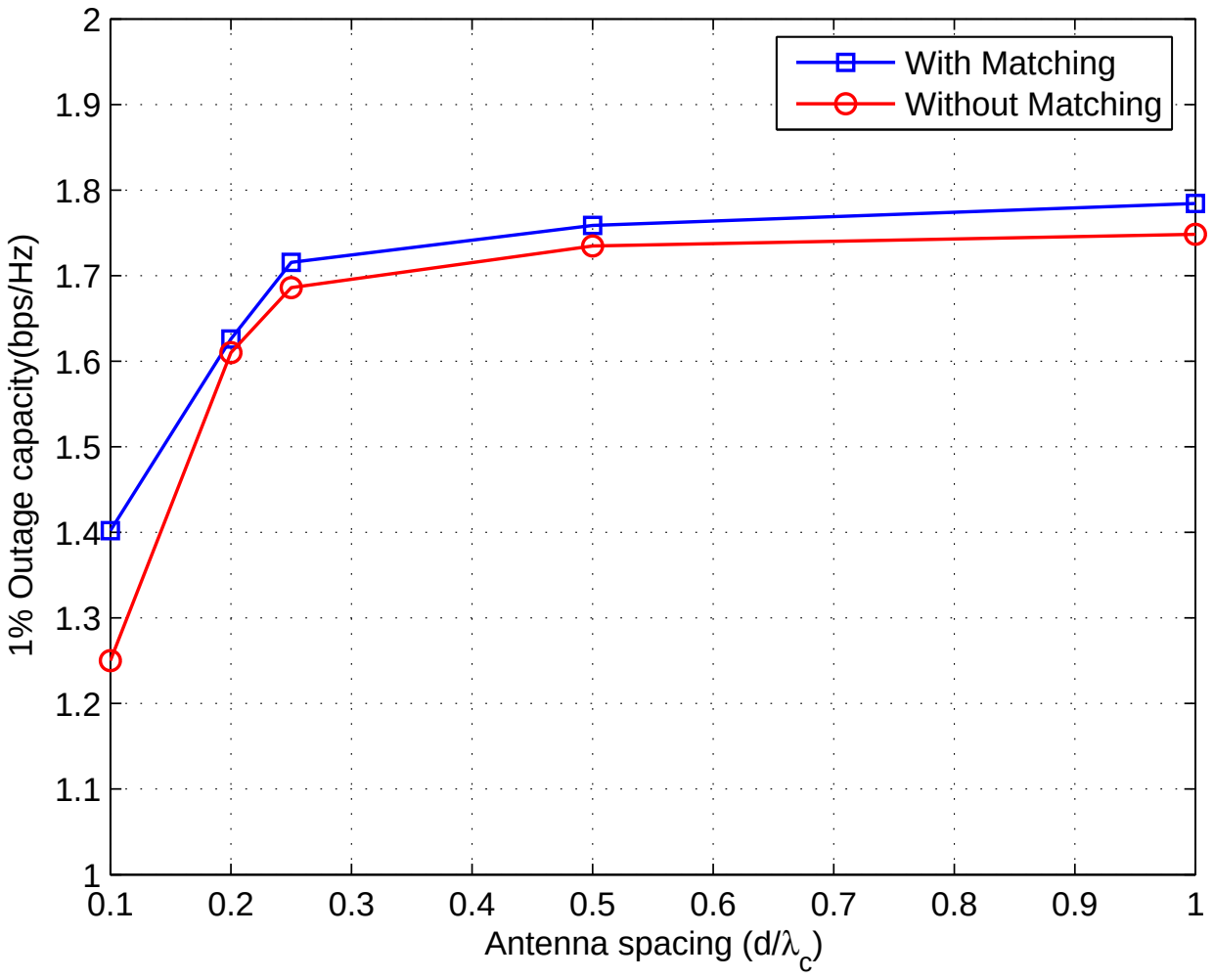

Fig. 11. Diversity-OFDM: Outage capacity vs. spacing, $N = 2$

It suffices to consider a pair of complex zeros in the right-half complex plane $z_{r1} = z_{r2}^*$ [17] (see Appendix), such that[10]

$$W \log\frac{1}{|\Gamma_0|^2} \leq \left(1 - W^2/4\right)\frac{2\pi}{Q} \ ,$$

and

$$|\Gamma_0|^2 = e^{-2\pi(1-W^2/4)/QW} \ .$$

The usable eigen-modes are determined based on the VSWR $< 2$ criteria. By adjusting the antenna lengths, the two eigen-mode resonant frequencies can be altered to maximize the overlap around $f_c$, so as to increase the diversity order of the system. The OFDM parameters for simulation are inspired by IEEE 802.11a standard, with $K = 64$ sub-carriers spanning a bandwidth of $B = 20$ MHz. The results are presented for a load-noise dominant scenario: $(T_A : T_f : T_r) = (1 : 2 : 0)$, for $W = 2\%$ and $\mathcal{E}_s/N_0 = 10$ dB SNR. We measure the system performance in terms of the outage capacity – probability that the capacity falls below a certain threshold $C_0$:

$$C_{out} = \Pr(C < C_0) \ .$$

Fig. 11 shows $1\%$ outage capacity for different antenna spacings for two element arrays. The performance predicted by these simulations shows that with increasing antenna spacing, the capacity increases for small spacings and starts to saturate around a quarter of a wave-length spacing. This behavior is in stark contrast to some of the studies which (assuming narrowband models) predict that, in theory, it is possible to achieve IID performance using optimal multiport matching networks even at very small spacings [7]. The answer clearly lies in bringing the bandwidth variable into the equation and re-visiting the communication-theoretic formulation for MIMO systems with mutual coupling. The results also suggest that in order to exploit diversity, practical arrays can be built with a separation of less than half-a-wavelength spacing – an oft-cited assumption in MIMO literature.

[10] Strictly speaking, considering only the first constraint yields an upper bound on capacity via $|\Gamma_0|^2 = e^{-2\pi/QW}$, while only the second constraint yields a lower bound via $|\Gamma_0|^2 = e^{-2\pi(1-W^2/4)/QW}$. For small $W$, the upper and lower bounds are quite tight.

Fig. 12 shows 1% outage capacity vs. antenna spacing for a four element UCA at $\mathcal{E}_s/N_0 = 10$ dB SNR, for the same settings. The impact of relative noise level from various noise sources, can be studied in a similar manner by varying $T_A$, $T_f$, and $T_r$ relative to $T_0$; the capacity still grows with increasing antenna separation and eventually saturates to i.i.d. case.

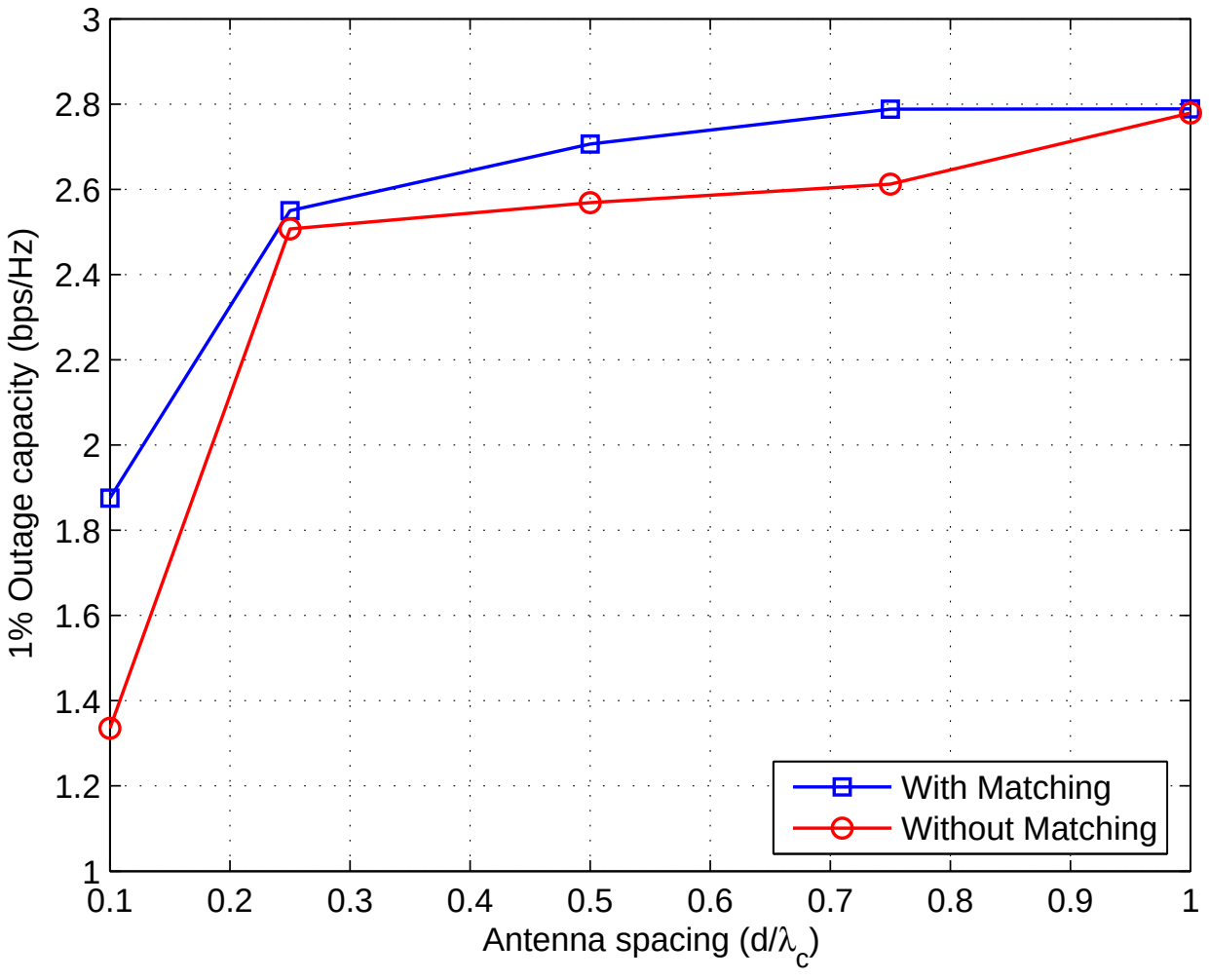

Fig. 12. Diversity-OFDM: Outage capacity vs. spacing, $N = 4$

## VII. Conclusion

In this paper, we primarily investigated the relationship between coupling and the bandwidth of compact multi-antenna systems, and derived the capacity limits of a diversity system with box-car matching. We analyzed the spatial modes of a compact antenna array and demonstrated that they exhibit different bandwidths and resonant frequencies; a phenomenon not observed in arrays with large antenna separation. Furthermore, we showed that traditional broadband matching theory can be applied to these eigen-modes in a straight-forward manner for uniform circular arrays. Although we considered equal bandwidths for all of the eigen-modes, a differentiated approach with respect to the mode bandwidths can be applied to further optimize the performance of Diversity-OFDM systems.

We discussed, how, in limit as the spacing between the antenna elements goes to zero, the most narrowband mode vanishes under strong coupling, leaving behind a system with a lower diversity order. The other key observation that the these modes are centered at different resonant frequencies – explains that a multiport matching network designed with narrowband assumptions is sub-optimal and that physically realizable optimal networks ought to be optimized over the bandwidth.

We also presented a communication-theoretic framework for Diversity-OFDM systems with mutual coupling and broadband matching. The results show that capacity increases with antenna spacing and that a quarter of a wavelength separation might suffice for most practical applications. An information-theoretic approach unifying the transceiver design with the overall system design, including signal processing aspects can lead to a new theory of compact MIMO communications.

## Appendix
## Matching Constraints

Let us consider a pair of complex zeros in the right-half complex plane $z_{r1} = z_{r2}^* = \alpha + j\beta$. Substituting normalized frequency $f_n = f/f_0$ in (11), the matching constraints are given by

$$\begin{aligned}
\int \log \frac{1}{|\Gamma(f)|^2}\, df &= \frac{\omega_0}{Q} - (z_{r1} + z_{r2}) \\
\int \log \frac{1}{|\Gamma(f)|^2}\, df &= \frac{2\pi f_0}{Q} - 2\alpha \\
\int \log \frac{1}{|\Gamma(f_n)|^2}\, df_n &= \frac{2\pi}{Q} - \frac{2\alpha}{f_0}
\end{aligned}$$

and

$$\begin{aligned}
\int \frac{1}{f^2} \log \frac{1}{|\Gamma(f)|^2}\, df &= \frac{4\pi^2}{\omega_0 Q} - \left(z_{r1}^{-1} + z_{r2}^{-1}\right) \\
\int \frac{1}{f^2} \log \frac{1}{|\Gamma(f)|^2}\, df &= \frac{2\pi}{f_0 Q} - \frac{2\alpha}{|z_{r1}|^2} \\
\int \frac{1}{f_n^2} \log \frac{1}{|\Gamma(f_n)|^2}\, df_n &= \frac{2\pi}{Q} - \frac{2\alpha f_0}{|z_{r1}|^2}
\end{aligned}$$

For a box-car reflection coefficient, $\Gamma(f_n) = \Gamma_0$ over the frequency-range

$$1 - W/2 \le f_n \le 1 + W/2\ ,$$

the two constraints yield:

$$\begin{aligned}
W G_0 &= \frac{2\pi}{Q} - \frac{2\alpha}{f_0} \\
\left(\frac{W}{1 - W^2/4}\right) G_0 &= \frac{2\pi}{Q} - \frac{2\alpha f_0}{|z_{r1}|^2}
\end{aligned}$$

where $G_0 \triangleq -\log|\Gamma_0|^2$. By choosing $|z_{r1}|^2 = f_0^2$, we get the least upper bound

$$\left(\frac{W}{1 - W^2/4}\right) G_0 \le \frac{2\pi}{Q} - \frac{2\alpha}{f_0}\ .$$

Choosing $\alpha \gg \beta$ (such that $\alpha \approx \sqrt{f_0}$ and $\alpha/f_0 \ll 1$), we have

$$G_0 \le \left(1 - W^2/4\right) \frac{2\pi}{QW}\ .$$

**Pawandeep S. Taluja** (S’10-M’11) was born in Kanpur, India, on January 22, 1982. In 2004, he received the B.Tech. degree in electrical engineering from the Indian Institute of Technology, Guwahati, India. He received the M.S. degree in electrical engineering as well as the Ph.D. degree in electrical engineering from North Carolina State University, Raleigh, NC in 2010 and 2011, respectively.

From 2004 to 2006, he worked at Samsung Electronics’ Wireless Terminal Division in Bangalore, India, with focus on UMTS protocol stack development. From 2006 to 2011, he was a graduate student working toward his Ph.D. in electrical engineering at North Carolina State University, Raleigh, NC. He is currently Staff Engineer, Communications Systems Group at MaxLinear, Carlsbad, California. His work focuses on broadband communication systems, specifically, digital front-end and modem design. His research interests lie in communication theory, signal processing, and information theory, including MIMO systems.

**Brian L. Hughes** (S’84-M’85) was born in Baltimore, MD, on July 16, 1958. In 1980, he received the B.A. degree in mathematics from the University of Maryland, Baltimore County. He received the M.A. degree in applied mathematics as well as the Ph.D. degree in electrical engineering from the University of Maryland, College Park, in 1983 and 1985, respectively.

From 1980 to 1983, he worked as a mathematician at the NASA Goddard Space Flight Center in Greenbelt, MD. From 1983 to 1985, he was a Fellow with the Information Technology Division of the Naval Research Laboratory in Washington, DC. From 1985 to 1997, he served as Assistant and then Associate Professor of Electrical and Computer Engineering at The Johns Hopkins University in Baltimore, MD. In 1997, he joined the faculty of North Carolina State University in Raleigh, where he is currently Professor of Electrical and Computer Engineering. His research interests include communication theory, information theory, and communication networks.

Dr. Hughes has served as Associate Editor for Detection of the IEEE Transactions on Information Theory, Editor for Theory and Systems of IEEE Transactions on Communications, and as Guest Editor for two special issues of IEEE Transactions on Signal Processing. He has also co-chaired the 2008 Globecom Wireless Communications Symposium, the 2004 Globecom Communication Theory Symposium, as well as the 1987 and 1995 Conferences on Information Sciences and Systems. He has also served on the program committees of numerous international conferences, including the IEEE Global Communications Conference, IEEE International Communications Conference, IEEE International Symposium on Information Theory, and the IEEE Wireless Communications and Networking Conference.